\documentclass[journal,twoside]{IEEEtran}
\usepackage[utf8]{inputenc}
\usepackage[T1]{fontenc}
\usepackage{amsmath,amssymb}
\usepackage{graphicx}
\usepackage{array}
\usepackage{cite}
\usepackage{textcomp}
\graphicspath{{./}}
\begin{document}

\title{Open-Set Vessel Re-Identification from Underwater Ship-Radiated Noise with a Raw-Waveform Selective-Kernel Acoustic Neural Network (SKANN) and a Cross-Passage Evaluation Protocol}

\author{Sunil~Tyagi%
\thanks{Preprint, September 2026.}%
\thanks{S. Tyagi is with Oravont Systems LLP, Noida, India (e-mail: styagi@oravontsystems.com; ORCID~0000-0001-5897-7955).}%
\thanks{An Indian provisional patent application (202611107132, filed 6 September 2026) covers aspects of the method described here.}%
}

\markboth{Preprint, September 2026}%
{Tyagi: Open-Set Vessel Re-Identification from Ship-Radiated Noise with SKANN}

\maketitle

\begin{abstract}
Underwater acoustic target recognition has converged on closed-set classification by vessel type, a task that does not answer the operational question of whether a monitoring system has heard \emph{this hull} before. We formalise open-set, cross-passage vessel re-identification on public hydrophone data and specify an evaluation protocol that removes the two easiest routes to a high score: hull-disjoint splits keyed to MMSI/IMO, galleries and queries drawn from disjoint passages of each hull, source-pure galleries, and an audio-adjudicated transit-deduplication gate. We describe SKANN, a raw-waveform encoder whose front end is a four-scale bank of learned filters (kernels of 16 ms to 1 s) fused by selective-kernel attention, trained with an additive angular-margin objective and an augmentation regime that perturbs recording chain, ambient noise and multipath while preserving the narrowband line structure that carries identity. On a 40-hull IARA gallery (96 queries against 98 passage candidates), cross-passage rank-1 is 0.25 for the embedding and 0.26 for a fully automated narrowband-tonal comparator; the two are statistically indistinguishable at the top of the ranking, the embedding orders the remainder of the list more reliably (AUC 0.82 vs 0.76), and their score fusion reaches rank-1 0.35 --- the only contrast that attains nominal significance, which we present as evidence of partial complementarity rather than as a recommendation. Transit deduplication alone removes a 16--21 point apparent rank-1 advantage, a larger effect than any between-method difference. Two further findings delimit what public data can support: ShipsEar cannot separate hull identity from recording channel under an identity protocol, and cross-network fine-tuning lifts performance on vessels seen during fine-tuning but is a null result on unseen ones. The results support analyst triage over a ranked shortlist, not identification. The checkpoint, validation embeddings, transit map and per-query outputs are released under CC-BY-4.0 (doi:10.5281/zenodo.22160138).
\end{abstract}


\section{Introduction}\label{sec:intro}

Passive acoustics is the natural sensing modality for wide-area maritime awareness: sound propagates efficiently underwater, hydrophones are cheap to operate continuously, and a vessel under way cannot avoid radiating the noise of its own machinery \cite{urick1983principles,nielsen1991sonar}. The machine-learning literature built on this modality --- underwater acoustic target recognition (UATR) --- has converged almost entirely on one problem template: \textbf{closed-set classification by vessel type}. Given a recording, predict \emph{cargo}, \emph{tanker}, \emph{ferry}, or \emph{background}, over a class list fixed at training time, on public corpora such as ShipsEar \cite{santosdominguez2016shipsear} and DeepShip \cite{irfan2021deepship}. On that template, reported accuracies on ShipsEar have reached the mid-90 \% range \cite{hummel2024survey,hong2021underwater}, and the benchmarks are widely regarded as approaching saturation \cite{hummel2024survey}.

The operational question, however, is usually a different one. A monitoring system that has previously heard a vessel should recognise \emph{that individual hull} when it passes again --- and should say ``unknown'' when a contact matches nothing it has heard before. This is \textbf{open-set re-identification}: identity, not category; a gallery that grows by enrolment rather than a class list frozen at training time; and a thresholded similarity decision rather than an argmax. It is the acoustic sibling of speaker verification \cite{snyder2018xvectors,desplanques2020ecapa} and face recognition \cite{schroff2015facenet,liu2017sphereface}, and it is nearly absent from the UATR literature, in part because the standard corpora and protocols were not designed for it.

The gap is not merely one of task definition; it is one of \emph{measurement}. Classification protocols routinely place segments of the same recording on both sides of the train/test split, and even identity-aware evaluations often score one half of a recording against its other half. Both choices reward the model for matching the \emph{recording} --- its channel, range, speed, and sea state --- rather than the \emph{vessel}. As we show, evaluation designed to remove these shortcuts yields a much harder and, we argue, much more honest picture of what current methods can do.

This paper makes three contributions.

\textbf{1. Problem formulation and protocol.} We formalise open-set, cross-passage vessel re-identification on public data: hull-disjoint train/validation splits keyed to vessel identity (MMSI/IMO); galleries and queries drawn from disjoint passages of each hull; source-pure galleries so cross-corpus hardware signatures cannot act as a ranking shortcut; and a transit-level deduplication gate that detects and merges recordings sampled from the same physical transit before any score is computed. On IARA the deduplication step alone removes a 16--21 point apparent rank-1 advantage created by scoring recordings of the same physical transit against each other, which suggests that some published results on public corpora are inflated by duplicate leakage.

\textbf{2. A raw-waveform selective-kernel encoder.} We describe SKANN,\footnote{Selective-Kernel Acoustic Neural Network. The acronym is also used by Shan et al. \cite{shan2025skann} for \emph{Selective Kernel Audio Neural Networks}, a closed-set classifier of mixed ship events on a constructed multi-vessel corpus. The two systems share the acronym and the selective-kernel fusion mechanism; they differ in input representation (raw waveform here), task (open-set identity retrieval here) and evaluation protocol. The name is retained here for continuity with the published checkpoint \cite{tyagi2026skann}.} an encoder that replaces the fixed spectrogram front end with a bank of dense learned 1-D filters at four kernel lengths spanning $\sim$16 ms to $\sim$1 s, fused per input by selective-kernel attention \cite{li2019selective} --- a fusion mechanism previously applied to underwater ship audio for closed-set event recognition \cite{shan2025skann}, here used on raw waveform as the front end of an identity embedding --- followed by a compact 2-D convolutional backbone and a 512-d L2-normalised embedding trained with an additive angular-margin (ArcFace) objective \cite{deng2019arcface}. Condition invariance is treated as a property of the \emph{data}, not the loss: an augmentation regime randomises recording-chain colouration, ambient noise, and multipath while provably preserving the narrowband line structure that carries hull identity; transforms that would move absolute line frequencies are deliberately excluded.

\textbf{3. Honest evaluation findings on public data.} Evaluated under the protocol of contribution 1, we report the results that the classification template does not surface: (i) cross-passage rank-1 on a 40-hull public gallery is 0.25 for the learned embedding and 0.26 for an automated narrowband-tonal baseline --- far below what closed-set numbers suggest --- and paired tests confirm the two methods are indistinguishable at the top of the ranking; the embedding ranks candidates more reliably below rank 1 (AUC 0.82 vs 0.76), a direction we report as the observed pattern rather than an established ordering, since the paired difference interval reaches zero; the methods' per-query genuine scores correlate at only 0.55, and their score fusion lifts rank-1 to 0.35 --- the one contrast that reaches nominal significance --- which we present as evidence of partial complementarity, not as an operational recommendation. Read together, these numbers support analyst triage over a ranked shortlist; they do not support treating either system as an identifier of record. (ii) ShipsEar, the field's most-used corpus, cannot separate hull identity from recording channel and operating state under an identity protocol --- on its single recording chain, same-vessel similarities fall \emph{below} different-vessel similarities --- and any mixed-corpus retrieval number in which it participates is at least partly a corpus-recognition number; and (iii) cross-network fine-tuning on a second hydrophone network lifts performance on vessels seen during fine-tuning but is a null result on unseen vessels --- a generalisation limit we attribute to sparse cross-encounter positive pairs in public data rather than to the loss or the architecture.

We regard (ii) and (iii) as findings of equal standing with the method itself: they delimit what public data can currently support, and they are the results a practitioner needs before trusting any reported re-identification number.

The remainder of the paper is organised as follows. Section~\ref{sec:related} reviews related work. Section~\ref{sec:problem} defines the task and evaluation protocol, including the transit-deduplication gate. Sections~\ref{sec:data}--\ref{sec:aug} describe the data, preprocessing, and the augmentation philosophy. Section~\ref{sec:arch} details the architecture and its interpretability properties; Section~\ref{sec:training} the training procedure. Section~\ref{sec:results} presents results on IARA, ShipsEar, and the cross-network experiment. Section~\ref{sec:discussion} discusses limitations, and Section~\ref{sec:conclusion} concludes. Model weights are archived with a DOI and SHA-256 for reproducibility (see the Data and Code Availability statement).

\section{Related Work}\label{sec:related}

\textbf{UATR as closed-set classification.} The machine-learning literature on ship-radiated noise --- underwater acoustic target recognition (UATR) --- is dominated by a single template: convolutional classifiers over Mel, CQT or LOFAR spectrograms, trained and tested on public corpora such as ShipsEar \cite{santosdominguez2016shipsear}, DeepShip \cite{irfan2021deepship} and, more recently, IARA \cite{dasilva2025iara} and Oceanship \cite{li2024oceanship}, with the label set fixed at training time. Reported accuracies on ShipsEar have reached the mid-90 \% range \cite{hummel2024survey,hong2021underwater}; the DeepShip dataset paper's own baseline is considerably lower \cite{irfan2021deepship}, and the survey of Hummel et al. \cite{hummel2024survey} gives the breadth of the field, which we do not attempt to enumerate. What these results share is the question they answer --- what \emph{type} of vessel --- and the protocols under which they answer it, which permit segments of the same recording to fall on both sides of the train/test split. None of them addresses \emph{which hull}, and none is designed to measure it.

\textbf{Learned and parameterised front ends.} A second line of work replaces the fixed spectrogram front end with a learned one. SincNet parameterises each first-layer filter as a band-pass with learned cut-offs \cite{ravanelli2018sincnet}; LEAF learns a Gabor filterbank with per-channel compression and normalisation \cite{zeghidour2021leaf}; and very deep raw-waveform CNNs dispense with a parameterised family altogether \cite{dai2017verydeep}. Closest to the present work, learnable filterbanks have been applied to maritime vessel classification \cite{elsborg2025acoustic}, sharing our motivation but targeting type classification rather than open-set identity. SKANN differs from all of these in three deliberate choices: dense, unconstrained kernels rather than a parameterised family; an explicitly long-kernel ladder, up to about one second, chosen to resolve the sub-2 kHz tonal band; and selective-kernel attention \cite{li2019selective} as the mechanism that fuses the scales per input. Selective-kernel fusion has itself been applied to underwater ship noise, for mixed-event recognition on a constructed multi-vessel corpus \cite{shan2025skann}; the present work differs in input (raw waveform, long-kernel ladder), objective (an identity embedding rather than an event class) and protocol (open-set, cross-passage), and our architectural claim is confined to that combination. A staged variant --- Gabor-constrained early training released to free kernels late --- connects naturally to \cite{zeghidour2021leaf} and is left as future work (\S\ref{sec:interp}).

\textbf{Metric learning and open-set verification.} The machinery for enrol-and-verify recognition is mature elsewhere. Contrastive and triplet objectives \cite{schroff2015facenet} gave way to angular-margin softmax losses --- SphereFace, CosFace and ArcFace \cite{liu2017sphereface,wang2018cosface,deng2019arcface} --- that shape an embedding space in which cosine similarity is the decision statistic. Speaker verification is the closest operational analogue: x-vectors \cite{snyder2018xvectors}, ECAPA-TDNN \cite{desplanques2020ecapa}, end-to-end verification losses \cite{wan2018generalized} and the VoxCeleb protocols \cite{nagrani2017voxceleb} together define how identity embeddings are trained, enrolled and thresholded on unseen speakers, and open-set recognition theory \cite{scheirer2013toward} formalises the reject option. The nearest procedural analogue to our setting is open-set speaker identification by scoring against enrolled embeddings \cite{affek2022openset}. What is missing in the underwater domain is therefore not a loss function; it is corpora and protocols that permit cross-passage identity evaluation at all.

\textbf{Classical tonal analysis.} Passive sonar analysis supplies the incumbent identity-bearing representations. LOFAR narrowband analysis and DEMON envelope-modulation analysis \cite{nielsen1991sonar,chen2021lofarenhancement,hashmi2022demon} expose machinery lines whose frequencies are tied to shaft and blade rate, from which shaft count, speed and course can be recovered \cite{hashmi2022demon}; they are physically interpretable, and they remain the input to current classifiers \cite{liu2024swin}. They are also brittle: Doppler shifts the absolute frequencies on which they depend, and line detection is governed by SNR and integration time \cite{chen2021lofarenhancement}. From this tradition descends template matching --- distance to a stored LOFAR template, lately learned as a Siamese similarity over sub-bands \cite{lu2022lofar} --- and it is the lineage to which our comparator belongs: an automated tonal line-matching method, evaluated under the same protocol as the embedding. Every comparison in this paper is automated against automated; we make no claim relative to analyst-in-the-loop performance.

\textbf{Individual-vessel identification: the prior art and the gap.} The idea that each hull has a unique signature is old \cite{eberlin1988underwater}, and the literature that acts on it supplies, separately, most of the ingredients this paper combines. \emph{Historical fingerprinting}: spectral acoustic-fingerprint recognition of individual noise sources \cite{zurek2014spectral}, and landmark fingerprinting of recurring time--frequency onsets \cite{hashmi2016landmark}. \emph{Passage-level temporal modelling}: hidden Markov models over vessel passages in a fixed region \cite{vieira2020underwater}. \emph{Open-class detection}: individual ship detection posed explicitly as an open-class problem, with impostor ships absent from training and speaker-recognition machinery (neural and i-vector/PLDA) brought across \cite{karakos2018individual} --- the closest precedent for the open-set half of our protocol. \emph{Registration-template matching}: few-shot identification by distance to enrolled templates under contrastive training, on a five-ship task \cite{nie2023contrastive}. \emph{Specific-ship identification data}: a corpus built for identifying particular ships within multi-target recordings, with classification-style baselines \cite{du2024qiandaoear22}. \emph{Learned individual identification}: closed-set identification over a fixed list of 100 hulls under a random sample-level split, most recently on microcontroller-class hardware \cite{altaf2026shipnn}; identity features designed against inter-identity similarity and intra-identity variation with speed and area \cite{kim2025vessel}; and generalisation to unseen individual vessels studied directly \cite{fang2025momentum}. \emph{Open-set ship recognition}: template matching with a rejection threshold over deep auditory features \cite{zheng2024opensetship}. Specific-vessel identification from optical satellite imagery \cite{bostaninezhad2026selfsupervised} is the cross-modality analogue and does not bear on the acoustic problem.

Prior work has thus established several ingredients of individual-vessel acoustic recognition --- open-class detection, registration-template matching, few-shot identification, learned vessel-specific features --- but we found no study that evaluates the full combination of open-set cross-passage re-identification, hull-disjoint splits, source-pure galleries and transit-level duplicate auditing on public data. That combination, and the negative results that come with it, is what this paper contributes. The enrol-and-verify machinery is mature in faces and voices, and the underwater domain has prototypes of open-class and registration-based recognition; what it lacks is a shared, leakage-resistant benchmark for independent-passage re-identification, and the corpora to build one.

\section{Problem formulation and evaluation protocol}\label{sec:problem}

\subsection{Re-identification, not classification}\label{sec:reid}

We address open-set acoustic vessel \textbf{re-identification}: given a passive recording of an unknown contact, determine \emph{which individual hull} it is --- or that it is not any hull previously observed. This differs from the classification protocol that dominates the underwater acoustic target recognition (UATR) literature \cite{hummel2024survey,hong2021underwater} in three ways that change both the architecture and the evaluation:

\begin{enumerate}

\item \textbf{Identity, not category.} The system must separate two different cargo ships, not merely label both ``cargo.'' The label space is the set of individual hulls, keyed by MMSI/IMO.

\item \textbf{Open set.} The correct answer may be ``not in the gallery.'' Query identities need not appear in training; the decision is a thresholded similarity, not an argmax over a fixed class list. Architecturally the problem is the same family as speaker and face verification \cite{snyder2018xvectors,deng2019arcface}.

\item \textbf{Enrolment without retraining.} Model weights are frozen after training; new vessels are added by a single forward pass that appends their embedding to a gallery. A classification head cannot do this --- its class list is fixed at training time.

\end{enumerate}

Formally: an encoder $f_\theta$ maps a passage (the renormalised mean of its segment embeddings, \S\ref{sec:embed}) to a unit vector $e \in \mathbb{S}^{511}$. A \textbf{gallery} $G = \{(e_j, y_j)\}$ enrols every retained passage individually with its hull label $y_j$; it is not collapsed to one embedding per hull. For a query $q$ the gallery is ranked by cosine $\langle q, e_j \rangle$ over passages. The reported rank of the true hull is the rank of its best-scoring same-hull passage, so a hull with several enrolled passages is credited once, at its best entry. The open-set score distributions (\S\ref{sec:metrics}) are formed from each query's per-hull maximum on the genuine side and every individual impostor passage on the impostor side. The output is the ranked passage shortlist together with the top score for the accept/reject decision.

\subsection{Cross-passage protocol}\label{sec:crosspassage}

The evaluation unit is the \textbf{passage} (one recording/encounter of a vessel). The protocol is \emph{cross-passage}: the gallery and query sets for a given hull are disjoint passages, so the model is never allowed to match a recording to itself or to a temporally overlapping capture. Within-passage or split-half protocols --- scoring the second half of a clip against its first half --- are reported in parts of the literature but overstate performance, because the two halves share channel, range, speed, and sea state; we regard them as smoke tests and do not use them here.

Galleries are \textbf{source-pure}: a query is scored only against gallery entries from the same corpus, so cross-corpus hardware signatures cannot leak into the ranking as a shortcut.

\subsection{Transit-level deduplication (data-integrity gate)}\label{sec:dedup}

Public hydrophone corpora can contain near- and exact-duplicate recordings: the same capture filed under two recording IDs, or two files sampled from a single continuous transit. IARA catalogues transits by recording ID with no station or timestamp field, so one physical transit can be catalogued under several IDs. Under a cross-passage protocol duplicates are corrosive in both directions --- a duplicated \emph{query--gallery} pair inflates scores (the model matches the capture, not the hull), while a single capture filed under two hull IDs is label noise and deflates them. We therefore treated the audit as a gate: no reported number until the evaluation corpus had been adjudicated from audio.

\textbf{Why metadata cannot adjudicate.} A first, metadata-only pass grouped recordings by a composite key (ship ID, duration, CPA time, environment). It showed that the key collides by construction: a 300 s recording with CPA at 150 s is a protocol default across more than 430 IARA recordings, leaving only 49 metadata groups with a distinctive configuration. The pipeline's division of labour follows from that finding --- metadata \emph{nominates} candidate pairs, audio \emph{adjudicates} them.

\textbf{Audio adjudication.} Before pairing, 308 recordings whose vessel identifier is non-numeric (unidentified contacts, which never enter the train or validation identity sets) were excluded, leaving 1,517 identified recordings for audit. For each, a residual spectrum was formed over a 120 s window with tonal lines and the broadband closest-point-of-approach envelope removed, isolating capture-specific \emph{texture} rather than vessel- or geometry-specific content. Texture correlation was computed over 19,238 pairs (all 15,723 same-hull pairs plus 3,515 cross-hull pairs nominated by metadata key or rare duration), and waveform cross-correlation over every pair above a low texture gate. The decision rule is joint --- same transit if texture correlation $\geq$ 0.40, or if texture $\geq$ 0.20 \emph{and} waveform correlation $\geq$ 0.15; same capture if waveform correlation $\geq$ 0.80 --- and was calibrated on eleven distinctive- configuration pairs known to be the same transit (11/11 recovered) rather than assumed. Both thresholds were checked against empirical cross-hull nulls: the texture bar sits above the 99th percentile of 300 cross-hull pairs, and the waveform bar about four times above the 95th percentile of 200 cross-hull non-candidates. A single texture cut high enough to clear the null cannot reach genuine pairs in the 0.20--0.37 band; those are recoverable only with waveform corroboration. Transits are the union of \emph{same-hull} confirmed edges only; cross-hull confirmations are reported as label-noise findings and never merged.

\textbf{Run history.} The adjudicator refused its own first configuration: a single texture cut at 0.10 was rejected by the script's built-in null check before any table was read. A second, uncalibrated single cut at 0.37 cleared the null but merged cross-hull edges and under-merged genuine pairs; it is superseded. The calibrated joint rule was validated against an acceptance test written a month earlier from an independent hand computation over the same pair measurements --- validation count, two named pair decisions, the train-side redundancy and illusory-hull counts, and the resulting table --- and reproduced it end-to-end from source audio. This is independent confirmation, not agreement by construction.

\textbf{Outcome.} The audit merged the training partition's 1,402 recordings into 1,250 transits (152 redundant, $\approx$11\%) and the validation partition's 115 recordings into \textbf{99 transits} over 40 hulls (16 redundant). Of 128 multi-recording transits, 128 were found by both metadata and audio, 72 by audio only, and 70 metadata nominations were rejected by audio: metadata-keyed deduplication both over- and under-merges. Thirty-one training hulls with two recordings collapse to a single transit, so their apparent multi-encounter status is illusory. Fifteen cross-hull pairs were flagged and none merged; they include one byte-identical file pair, one co-transit scene filed under two identities, and one cluster of twelve contiguous recording IDs across many hulls with high shared texture but near-zero waveform correlation --- a shared site/session soundscape, not shared audio, and the concrete reason the waveform check is load-bearing. Every cross-hull finding is train/train; the validation partition is untouched by all of them.

All reported metrics are computed at the \textbf{transit level} under the adjudicated map --- each retained transit contributes one query and one gallery entry --- so that no same-capture pair can appear on both sides of a comparison. We report the procedure in full because earlier evaluation passes on the raw recording list produced materially different numbers (inflated by within-hull duplicates; in the over-merged intermediate pass, deflated); only the audio-adjudicated map is used for every number in this paper.

\subsection{Metrics}\label{sec:metrics}

For each query transit with its true hull present in the gallery:

\begin{itemize}

\item \textbf{rank-1} --- fraction of queries whose top-ranked gallery vessel is correct;

\item \textbf{hit@k} --- fraction with the correct vessel in the top \emph{k};

\item \textbf{median rank (medR)} --- median rank of the true vessel;

\item \textbf{AUC} --- area under the ROC over the pooled same-vessel and different-vessel score distributions, i.e. the probability that a random genuine comparison out-scores a random impostor comparison. This is the open-set quantity: it measures the score separability on which any accept/reject threshold operates;

\item \textbf{EER} --- the equal-error rate of the same distributions, used during training as a compact tracking scalar.

\end{itemize}

No classification metrics (accuracy, F1, confusion matrices) are reported; they do not apply to open-set re-identification.

\textbf{Sampling unit and inference.} Because leave-one-passage-out queries from the same hull share gallery structure, intervals and paired tests treat the \textbf{hull}, not the query, as the sampling unit: 95\% intervals on AUC and median rank are hull-cluster bootstrap percentiles (B = 10 000), and between-method differences on hit@k are tested with McNemar's exact test on the discordant queries. Rank-1 and hit@k carry no interval of their own; the paired tests are the inferential statement for those columns. Contrasts are reported uncorrected, with the number of contrasts stated.

\textbf{Comparator and score fusion.} The comparator throughout is a fully automated narrowband-tonal method (TPSW whitening, line extraction, matched-line-fraction scoring) --- automated versus automated; no analyst-in-the-loop numbers are reported. Fusion is a \textbf{per-query gallery-referenced z-score}: for each query, each method's similarity row is z-scored against all non-self gallery entries (population standard deviation) and the two z-rows are averaged. Fusion is reported as an experiment alongside the two methods, not as a recommendation.

\section{Data}\label{sec:data}

Three public corpora are used, in two roles (Table~\ref{tab:corpora}). All are used under their owners' terms and cited rather than redistributed; the deposit contains no audio and no derived tensors. The IARA snapshot used throughout is Zenodo record doi:10.5281/zenodo.15758636, byte-identical to record 15777429, which the IARA paper cites, the two differing only in a summary figure. The published checkpoint, the validation passage embeddings, the audio-adjudicated transit map of \S\ref{sec:dedup} and the script that produced it, and the similarity matrices and per-query results behind \S\ref{sec:iara} are deposited under CC-BY-4.0 at doi:10.5281/zenodo.22160138.

\begin{table}[!t]
\caption{Corpora and roles.}
\label{tab:corpora}
\centering
\footnotesize
\begin{tabular}{>{\raggedright\arraybackslash}p{2.5cm}>{\raggedright\arraybackslash}p{3.0cm}>{\raggedright\arraybackslash}p{1.9cm}}
\hline
Corpus & Role & Identity labels \\
\hline
IARA \cite{dasilva2025iara} & Base training + primary validation & MMSI/IMO \\
ShipsEar \cite{santosdominguez2016shipsear} & Base training + validation & Per-vessel \\
ONC (Ocean Networks Canada) \cite{onc2020sog} & Cross-network fine-tuning experiment & MMSI (AIS-derived) \\
\hline
\end{tabular}
\end{table}

\textbf{Identity keying.} Every recording is keyed to an individual hull via MMSI/IMO where available. The identity set is deliberately singleton-heavy --- of 646 training hulls, 500 have a single recording --- which is representative of what public corpora actually supply and directly shapes two training choices (the ArcFace margin, \S\ref{sec:objective}, and the epoch-set resampling, \S\ref{sec:epochset}).

\textbf{Hull-disjoint split, frozen before segmentation.} Training and validation are split \emph{by hull}, not by clip fraction: 646 training hulls and 49 validation hulls (40 IARA, 115 recordings; 9 ShipsEar, 37 recordings) with zero overlap. The split is decided and written to file \textbf{before} segmentation or augmentation runs, so a validation hull cannot leak into training through a segment or an augmented copy. Every validation hull has at least two recordings of verifiable identity; whether those recordings constitute independent passages is decided by the audit of \S\ref{sec:dedup} (IARA) or by the corpus's own session structure (ShipsEar). After deduplication three IARA hulls retain a single transit, and four of the nine ShipsEar hulls have a single recording session (their several recordings are docking-state repeats --- arriving, manoeuvring, waiting, departing), leaving 37 IARA hulls and 5 ShipsEar hulls able to supply queries; ShipsEar's 37 recordings yield 14 scoreable passages. Hull counts and the 500-single-recording figure above are at the recording level as used in training; at the transit level the audit adds 31 further training hulls whose two recordings are one transit. After preprocessing, the corpus comprises 322,159 segment tensors (314,520 train / 7,639 validation).

\textbf{ONC as the cross-network instrument.} ONC is an independent hydrophone observatory, distinct in hardware and deployment from IARA/ShipsEar, and is used to ask the transfer question: does a fingerprint learned on one recording network survive on another? A multi-passage ONC subset is split into enrolled and held-out vessels; the held-out vessels are never trained on at any stage and provide the unseen cross-network verdict (\S\ref{sec:onc}). ONC identities were assigned from AIS: recordings are 300 s files from the SCVIP icListen hydrophone in the Strait of Georgia, and a file was keyed to a vessel's MMSI when exactly one AIS-broadcasting vessel, as received by ONC's own AIS station, lay within 2 km of the hydrophone during the window and no other AIS-broadcasting vessel lay within 6 km; files that failed either condition were not used.

\section{Preprocessing}\label{sec:preproc}

The guiding principle is to \textbf{strip recording-chain artefacts} so the encoder keys on vessel identity rather than on which hydrophone or channel made the recording. Three decisions implement it, applied identically to every corpus.

\textbf{Resample to 8 kHz mono.} The discriminating tonal structure of surface vessels sits below $\sim$2 kHz (shaft and blade-rate harmonics, auxiliary machinery lines) \cite{urick1983principles}. Sampling at 8 kHz preserves that band with a comfortable anti-alias margin (Nyquist 4 kHz) while discarding high-frequency content that carries little identity, and standardizes all corpora to a common footing.

\textbf{Segment into 5 s / 40,000-sample windows.} Each recording is cut into non-overlapping 5-second windows, the canonical tensor being $(1, 1, 40000)$ float32. Five seconds is long enough to resolve low-frequency lines ($\sim$1 Hz resolution is reachable by the longest filterbank kernel, \S\ref{sec:filterbank}) and short enough to yield many segments per passage. A leftover tail shorter than 1 s is dropped; a tail of at least 1 s is captured by taking the last 5 s from the end of the recording.

\textbf{Per-segment z-normalisation.} Each segment is normalised to zero mean and unit variance in the time domain. This removes absolute level --- a recording-side nuisance carrying no identity --- and is complementary to, not redundant with, the L2-normalisation of the output embedding: the input z-norm removes level from the \emph{signal}; the output L2-norm removes magnitude from the \emph{embedding}, which is what makes cosine the correct similarity.

\section{Augmentation}\label{sec:aug}

Augmentation defines what the embedding is \textbf{invariant to}: each transform perturbs one physical nuisance variable that changes between recordings of the same hull, while preserving the identity cue, so that gradient descent is forced onto the invariant that remains. All augmentation is pre-computed during preprocessing and applied to training hulls only; validation is never augmented. Each training segment yields one original plus three augmented copies, each copy a random combination of several transforms.

The five transforms, at the level of the physical variable each simulates:

\begin{enumerate}

\item \textbf{Random spectral tilt (EQ).} A gentle random re-colouring of the spectrum, with the identity band tilted only mildly and low-frequency attenuation capped so a hydrophone's roll-off is never deepened. Simulates recording-chain frequency response --- the headline defence against the model shortcutting on hardware signature rather than hull.

\item \textbf{Ambient mixing.} Real ambient noise from a curated bank is added at a randomised signal-to-noise ratio. Simulates range, sea state, and background regime.

\item \textbf{Synthetic multipath.} A sparse randomised channel impulse response (direct path plus a small number of attenuated reflections, with surface bounces sign-flipped and total reflected energy capped) is convolved with the segment. Simulates boundary reflections. Critically, multipath moves line \emph{amplitudes}, never line \emph{frequencies}: a before/after peak-frequency check is logged for every augmented copy to verify the identity cue is preserved.

\item \textbf{Circular time shift.} A random roll of the segment. Simulates arbitrary event arrival time within the window.

\item \textbf{Ambient-fill masking.} A short region of the segment is replaced with low-level ambient noise --- never zeros, because a hard-zero gap creates a discontinuity the model would learn as a spurious feature. Simulates brief signal loss.

\end{enumerate}

\textbf{Deliberately excluded}, with reasons that are themselves informative:

\begin{itemize}

\item \textbf{Global gain scaling} --- provably inert: per-segment z-normalisation is scale-invariant and every transform above is scale-equivariant, so gain rides through the chain and is cancelled before the tensor is saved.

\item \textbf{Doppler/resample warping} --- corrupts identity: warping the time axis shifts the \emph{absolute} tonal frequencies, which \emph{are} the identity cue.

\item \textbf{Phase randomisation} --- corrupts identity: destroys the temporal structure the fingerprint depends on.

\end{itemize}

The exclusion list is where this augmentation philosophy differs most sharply from speech and general-audio practice, where pitch/speed perturbation is a staple: for hull identity, the frequencies are the payload.

\section{Architecture}\label{sec:arch}

\begin{figure*}[!t]
\centering
\includegraphics[width=\textwidth]{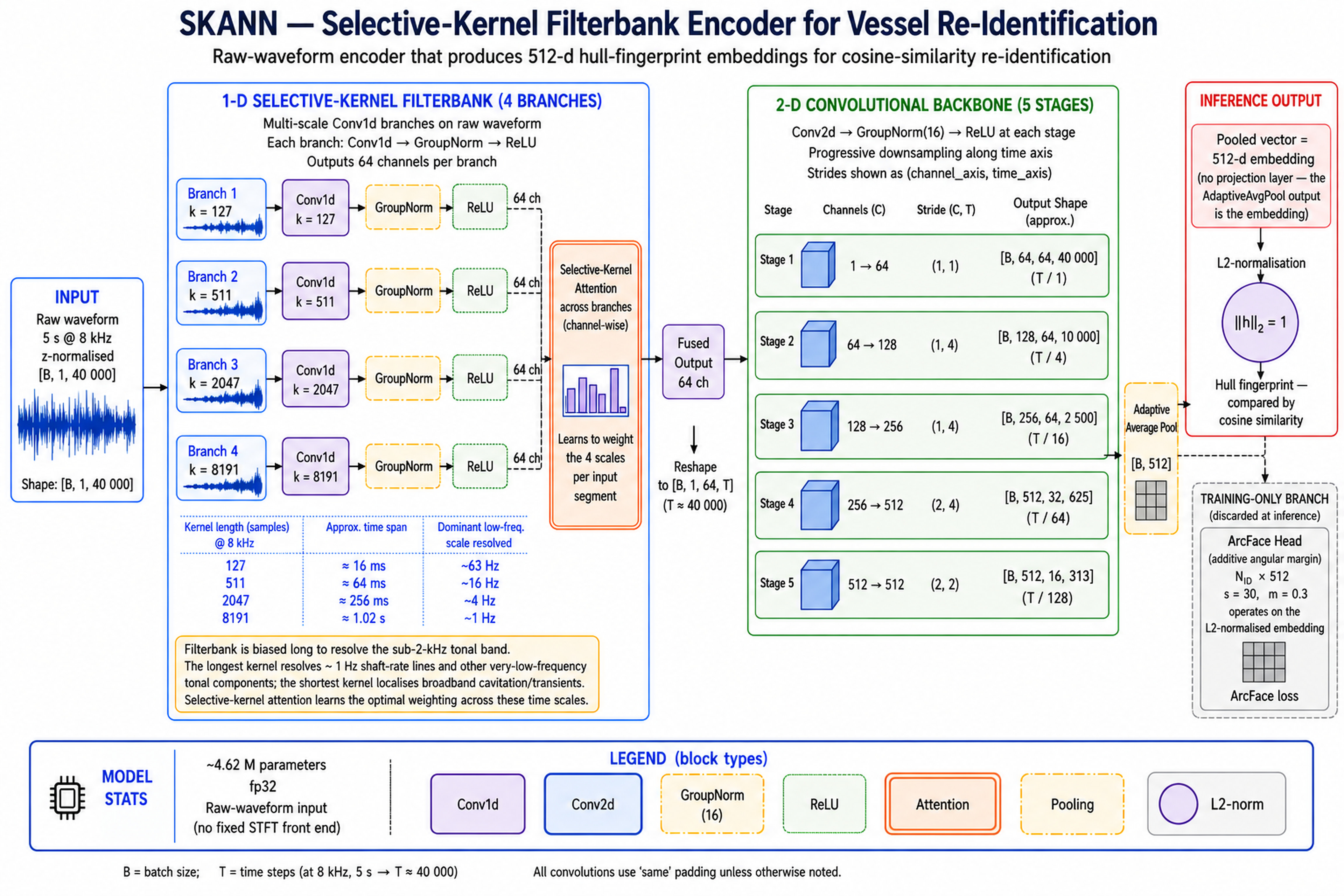}
\caption{SKANN architecture. A 5 s raw-waveform segment (z-normalised, 8 kHz) enters a four-branch 1-D filterbank with learned kernels of 127, 511, 2047 and 8191 samples ($\approx$16 ms to 1 s), 64 channels per branch; selective-kernel attention weights the four time scales channel-wise for each segment. A five-stage 2-D convolutional backbone (1$\rightarrow$64$\rightarrow$128$\rightarrow$256$\rightarrow$512$\rightarrow$512 channels) downsamples the time axis early and the learned spectral axis late; adaptive average pooling yields a 512-d L2-normalised embedding with no projection layer. There is no fixed STFT or mel front end. The ArcFace head is used in training only; at inference, embeddings are compared by cosine similarity.}
\label{fig:arch}
\end{figure*}

SKANN is a raw-waveform convolutional encoder, $\sim$4.62 M parameters, mapping a z-normalised 5 s segment $(B, 1, 40000)$ to a 512-d L2-normalised embedding. There is no fixed STFT or mel front end (Fig.~\ref{fig:arch}).

\subsection{Selective-kernel filterbank (learned multi-resolution front end)}\label{sec:filterbank}

Four parallel 1-D convolution branches run over the raw waveform with kernel lengths \textbf{127 / 511 / 2047 / 8191} samples, 64 channels each, each branch Conv1d $\rightarrow$ GroupNorm \cite{wu2018group} $\rightarrow$ ReLU with `same' padding. At 8 kHz the kernels span roughly 16 ms / 64 ms / 256 ms / 1.02 s, giving best-case frequency resolutions of roughly 63 / 16 / 4 / 1 Hz. The ladder is biased long to resolve the sub-2-kHz tonal band: a short kernel localises broadband transients in time; the longest kernel can resolve $\sim$1 Hz shaft-rate line spacing. Kernel \emph{length} fixes only resolution, not band --- every kernel's frequency response is defined over the full 0--4 kHz range, and where each tunes itself is left entirely to training (a point \S\ref{sec:interp} returns to).

A \textbf{selective-kernel attention} stage \cite{li2019selective} then fuses the branches: the four $(64, T)$ branch maps are summed and averaged over time to a 64-number channel summary, squeezed through a Linear(64$\rightarrow$4) bottleneck with ReLU, and expanded by four per-branch Linear(4$\rightarrow$64) layers whose outputs are softmaxed \emph{across branches} --- so for each of the 64 channels the four branch weights sum to one. The branch maps are blended under these weights into a single fused $(64, T)$ map. The network thereby decides, per segment and per channel, how much to trust each time scale; the mixing is input-dependent but involves no token-to-token interactions.

\subsection{2-D convolutional backbone}\label{sec:backbone}

The fused map is reinterpreted as a one-channel image of height 64 (the learned-filter axis, playing the role frequency plays in a spectrogram) and width $T$, so that 3$\times$3 kernels can learn cross-channel temporal co-occurrence patterns. Five blocks of [Conv2d 3$\times$3 $\rightarrow$ GroupNorm(16) $\rightarrow$ ReLU] follow, with channels \textbf{1$\rightarrow$64$\rightarrow$128$\rightarrow$256$\rightarrow$512$\rightarrow$512} and strides (h, w) = \textbf{(1,1), (1,4), (1,4), (2,4), (2,2)} (Table~\ref{tab:backbone}).

\begin{table}[!t]
\caption{Backbone stages: channels, strides and output shapes for a 5\,s input.}
\label{tab:backbone}
\centering
\footnotesize
\begin{tabular}{lccc}
\hline
stage & channels & stride (h,w) & output shape \\
\hline
input & 1 & --- & (B, 1, 64, 40000) \\
1 & 1$\rightarrow$64 & (1,1) & (B, 64, 64, 40000) \\
2 & 64$\rightarrow$128 & (1,4) & (B, 128, 64, 10000) \\
3 & 128$\rightarrow$256 & (1,4) & (B, 256, 64, 2500) \\
4 & 256$\rightarrow$512 & (2,4) & (B, 512, 32, 625) \\
5 & 512$\rightarrow$512 & (2,2) & (B, 512, 16, 313) \\
\hline
\end{tabular}
\end{table}

The design logic: time starts enormous and is collapsed early (width strides of 4); the 64-row channel axis is information-dense --- each row a distinct learned spectral view --- and is preserved through stage 3 and collapsed late; channel depth grows as spatial resolution is spent.

\subsection{Embedding and training head}\label{sec:embed}

Adaptive average pooling collapses the final $(512, 16, 313)$ map to a 512-d vector, which is \textbf{L2-normalised to the unit hypersphere; there is no projection layer --- the pooled vector is the embedding}. During training only, an ArcFace head (one learned prototype direction per training hull) sits on the normalised embedding for the loss; \textbf{at inference the head is discarded} and matching is plain cosine against the enrolled gallery. A passage's fingerprint is the renormalised mean of its segments' unit vectors.

\subsection{What the front end learns (interpretability)}\label{sec:interp}

\begin{figure}[!t]
\centering
\includegraphics[width=\columnwidth]{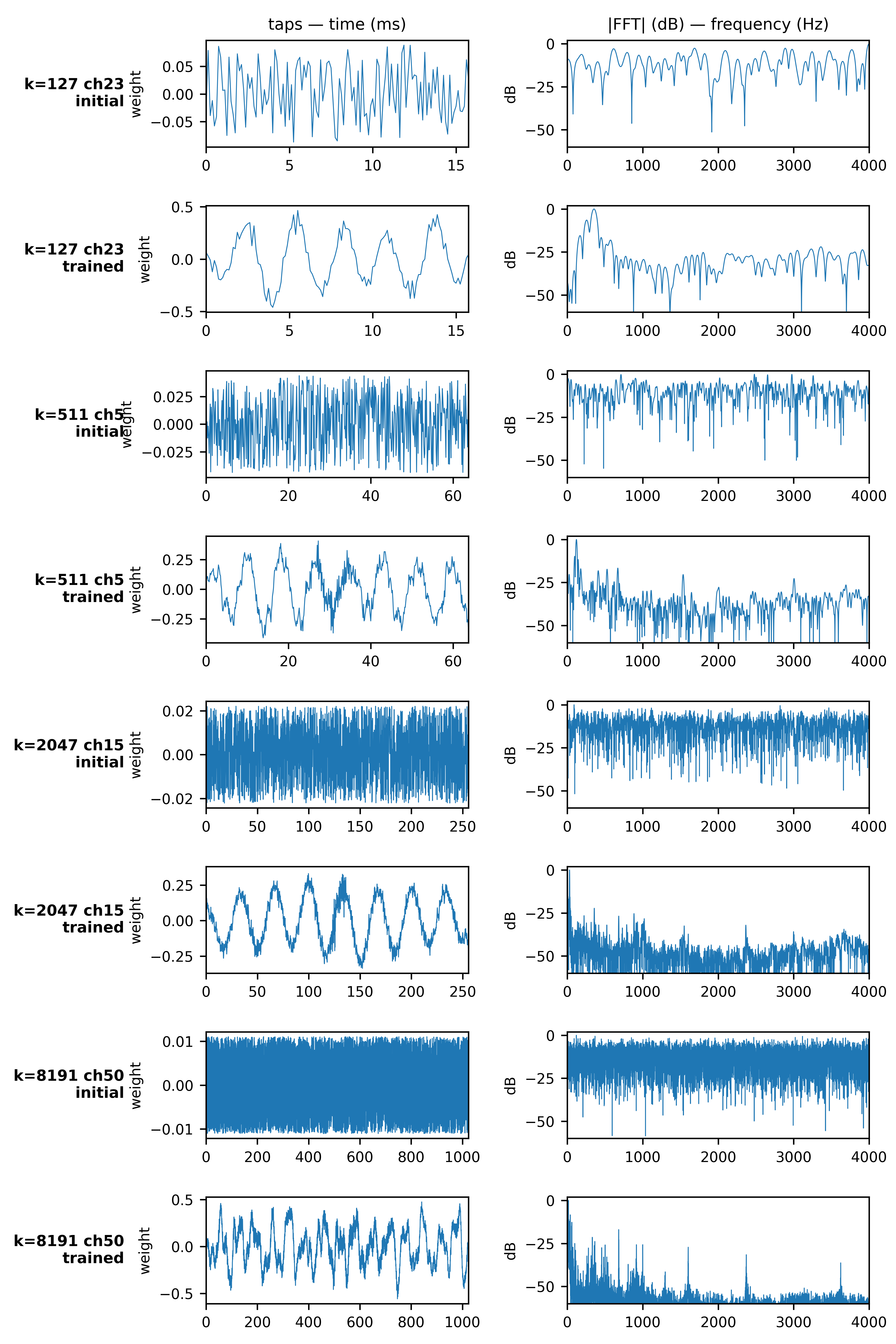}
\caption{Selective-kernel filterbank: initial versus trained kernels (epoch 21). For each branch, the trained kernel with the most spectrally peaked response is shown as taps and |FFT| (dB) beside its random initialisation. The three shorter branches converge to windowed sinusoids of roughly constant 6--8 cycles per window, peaking near 354, 121 and 30 Hz (k = 127, 511, 2047) --- a constant-Q-like decomposition that nothing in the architecture imposes; the longest branch (k = 8191) shows multiple discrete lines, its strongest at $\approx$15 Hz. The plotted kernels are selected for spectral peakedness, so narrowband specialists are shown by construction; across all 64 kernels per branch the median spectral crest factor rises monotonically with kernel length (4.5, 14.4, 38.7, 68.9). Time panels span 16 ms to 1.02 s (64$\times$), so oscillation density is not comparable across rows.}
\label{fig:kernels}
\end{figure}

Because the filterbank kernels are free tap vectors initialised from noise, their trained shapes are an empirical readout of what the task demanded. Inspecting the trained kernels --- for each branch, the kernel with the most spectrally peaked response (Fig.~\ref{fig:kernels}) --- shows two structures:

\textbf{Constant-Q emergence.} The clearest specialists of the three shorter branches are windowed sinusoids at a roughly constant 6--8 cycles per window: peak response at approximately 354 Hz (k = 127), 121 Hz (k = 511), and 30 Hz (k = 2047) --- centre frequency falling roughly in inverse proportion to kernel length. Nothing architectural forces this coupling: kernel length sets resolution ($\sim$SR/k), not band, and every branch is free to tune anywhere in 0--4 kHz. Training, from random initialisation, spontaneously reconstructed a constant-Q, wavelet-like decomposition. All specialists tuned below $\sim$500 Hz --- the vessel-tonal identity band --- which is the physics of the task showing through the optimisation.

\textbf{Harmonic-comb detectors.} The longest branch (k = 8191) grew kernels whose spectra show multiple discrete lines rather than a single peak --- its plotted exemplar's strongest line sits at $\approx$15 Hz, squarely in shaft-rate territory --- detectors matched to harmonic families (shaft/blade-rate stacks) as a whole. This is structure a single Morlet/Gabor atom cannot express, and it bears directly on the question a reviewer will ask: \emph{why not just use a CWT or a Gabor-parameterised front end?} The answer this inspection gives is that the free front end converges toward a constant-Q decomposition \emph{plus} comb detectors that no fixed wavelet family contains.

Two caveats accompany the figure. First, a \textbf{selection caveat}: the plotted kernels are each branch's most spectrally peaked, so narrowband specialists are shown by construction; the figure does not establish that all 64 kernels per branch converge similarly, and transient- and comb-shaped kernels have low spectral crest and are excluded by that selector. A branch-level statistic partially offsets this: the median spectral crest factor across each branch's 64 trained kernels rises monotonically with kernel length --- 4.5 / 14.4 / 38.7 / 68.9 for k = 127 / 511 / 2047 / 8191 --- so the drift toward line-structured responses on the longer branches is a property of the branch populations, not only of the plotted exemplars. Second, an \textbf{axis caveat}: the four time panels span 16 ms to 1.02 s --- a 64$\times$ range --- so apparent oscillation density is not comparable across panels; frequency should be judged from the spectral column only.

A natural next step, not attempted here, is a staged-Gabor front end --- Gabor-parameterised or -initialised early epochs, then releasing the taps for free training, dropping the learning rate $\sim$10$\times$ at the release and releasing the longest branch last, since that is where comb detectors live. Precedents: SincNet (Ravanelli \& Bengio, 2018) \cite{ravanelli2018sincnet}; LEAF (Zeghidour et al., 2021) \cite{zeghidour2021leaf}. No front-end experiments beyond the free filterbank are reported.

\section{Training}\label{sec:training}

\subsection{Objective}\label{sec:objective}

Training uses \textbf{ArcFace} \cite{deng2019arcface} --- an additive angular-margin classification loss over the 646 training identities --- with the pre-head 512-d embedding retained for matching. Scale s = 30 and margin m = 0.3. The margin is deliberately gentler than the common 0.5 because the identity set is singleton-heavy (500 of 646 training hulls have a single recording); an aggressive margin on thin identities hurts more than it helps.

Supervised metric learning is used precisely because identity labels exist; there is no self-supervision and no positive-pair construction. Invariance to speed, channel, and conditions is not assumed to come from the loss --- it comes from the data (the augmentation regime, \S\ref{sec:aug}). Loss and invariance are treated as separate levers.

\subsection{Optimisation}\label{sec:optim}

Physical batch 16, gradient-accumulated to an effective batch of 64 (ArcFace does not rely on in-batch negatives, so the physical batch is purely a memory/speed knob). AdamW \cite{loshchilov2019decoupled}, learning rate 1e-3 with a 3-epoch linear warmup from 1e-4 and cosine decay toward 1e-6; decoupled weight decay 1e-4; gradient-norm clipping at 10.

\subsection{Epoch-set data design}\label{sec:epochset}

Rather than a weighted sampler, each epoch draws from one of 25 pre-built epoch manifests under an \emph{originals-preferred} resampling policy, rotating through the manifests across epochs. Per identity and per epoch set: hulls with abundant original recordings contribute a fixed quota of originals and no augmented copies; mid-sized hulls contribute all their originals topped up with augmented copies to the quota (the copies cycling across sets); thin hulls contribute all of their original-plus-augmented tensors every set. The effect is that every real recording and every identity stays in play across training while per-epoch identity balance is maintained, and augmented copies are used only where originals run short. Loading within an epoch is a plain shuffle. (This replaced an earlier per-hull cap, which was found to silently discard a large fraction of real recordings.)

\subsection{Validation, model selection and reproducibility}\label{sec:validation}

Validation is passage-level and runs on original (never augmented) segments of the 49 hull-disjoint validation vessels: a passage's segment embeddings are mean-pooled and renormalised, and all cross-passage pairs are scored by cosine. The tracked quantity is the same-vessel vs different-vessel cosine gap, from which EER and rank-1 follow. At the recording/session level used during training (which predates the adjudication of \S\ref{sec:dedup}), this validation pool comprises 133 passage embeddings supplying 129 same-vessel pairs against 8,649 impostor pairs; because the genuine side is thin, the trend across epochs is read rather than any single epoch.

\begin{figure}[!t]
\centering
\includegraphics[width=\columnwidth]{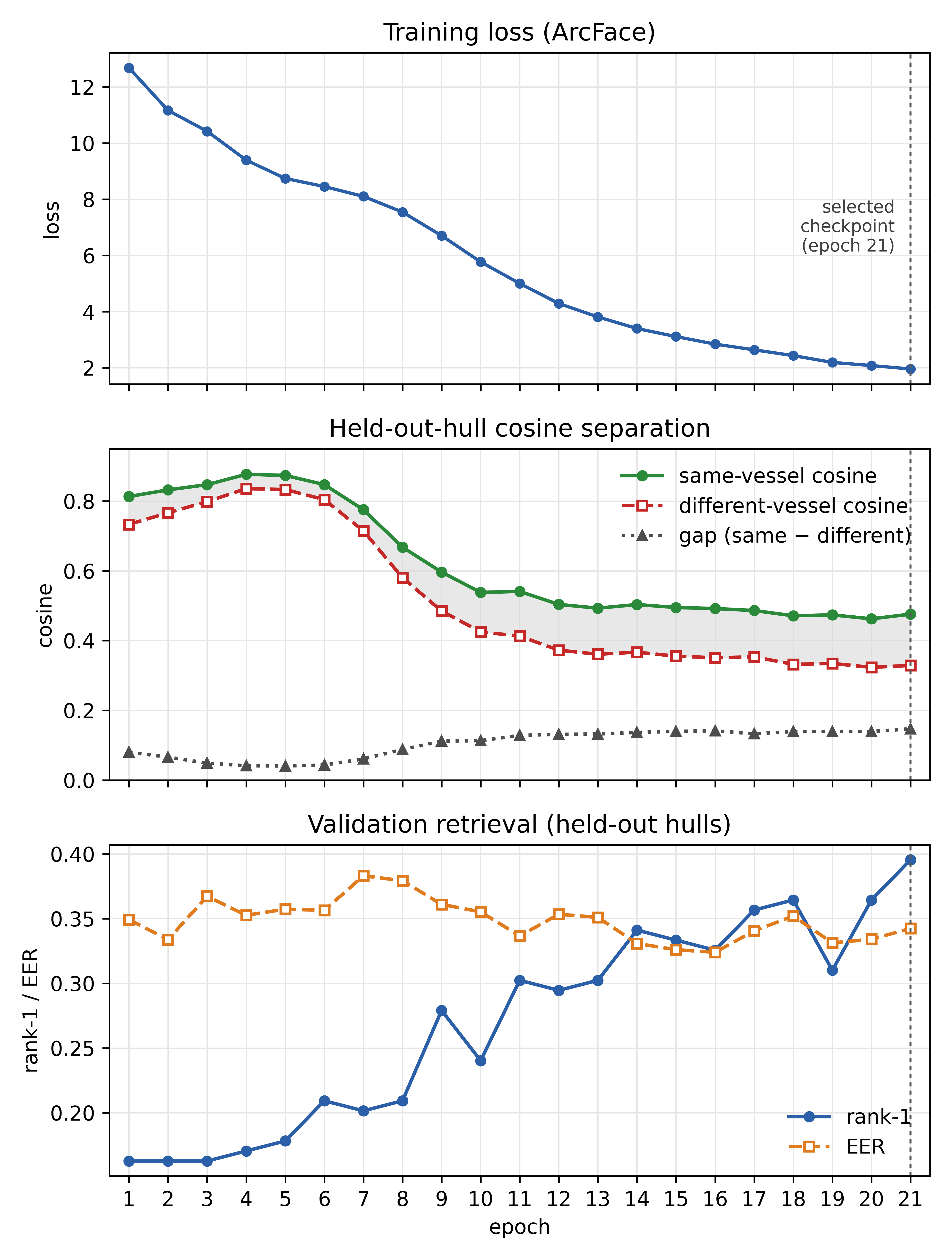}
\caption{Training dynamics and model selection. ArcFace loss, held-out-hull cosine separation (same-vessel, different-vessel, and their gap) and validation rank-1/EER by epoch, shown to the selected epoch 21; the run continued to epoch 27. Epoch 21, the running best by passage-level gap, is the frozen checkpoint evaluated throughout this paper and deposited at \cite{tyagi2026skann}. All curves are computed on the training-time validation pool of \S\ref{sec:validation} (recording-level, prior to the \S\ref{sec:dedup} transit adjudication) and are not comparable to the \S\ref{sec:iara} cross-passage figures.}
\label{fig:training}
\end{figure}

The run was configured for 40 epochs with early stopping at a patience of 10 on the validation gap. It was stopped after \textbf{27 epochs} by a compute-session limit; early stopping never triggered (the best epoch, 21, would have exhausted patience at epoch 31). Checkpoints were written every three epochs for the first nine and every epoch thereafter, together with a running best by passage-level gap. The checkpoint evaluated throughout this paper is the \textbf{best-validation epoch, 21} (Fig.~\ref{fig:training}), which is the checkpoint deposited at doi:10.5281/zenodo.22160138. Re-extracting the passage embeddings from that deposited checkpoint reproduces every reported similarity to within $10^{-4}$ (maximum absolute deviation 7.3 $\times$ $10^{-5}$ over the validation similarity matrix), and every reported table to four decimal places; the deposited passage vectors are the output of that re-extraction, and every table in \S\ref{sec:iara} can be regenerated from the deposited similarity matrices without audio or model.

\section{Results}\label{sec:results}

All numbers in this section are computed under the audio-adjudicated transit map of \S\ref{sec:dedup} with source-pure galleries and the comparator and fusion definitions of \S\ref{sec:metrics}.

\subsection{IARA cross-passage (primary result)}\label{sec:iara}

Evaluation is leave-one-passage-out. Deduplication reduced the validation corpus from 115 recordings to 99 transits over 40 hulls (3 $\times$ 1, 19 $\times$ 2, 14 $\times$ 3, 4 $\times$ 4 passages per hull). The three single-passage hulls cannot serve as queries, leaving \textbf{96 scoreable queries} drawn from 37 hulls, each ranked against the other \textbf{98 passage-level gallery entries}; galleries enrol every retained passage individually and are not collapsed to one embedding per hull.

Because leave-one-passage-out queries from the same hull share gallery structure, intervals and paired tests treat the \textbf{hull}, not the query, as the sampling unit: 95\% intervals on AUC and median rank are hull-cluster bootstrap percentiles (B = 10 000); between-method differences on hit@k are tested with McNemar's exact test on the discordant queries. One query is $\approx$ 1.04 percentage points. Results are given in Table~\ref{tab:iara}.

\begin{table*}[!t]
\caption{Cross-passage retrieval on IARA: each of the 96 queries (37 hulls) is ranked against the remaining 98 passage-level entries of the deduplicated 99-transit, 40-hull set. Bracketed intervals are 95\% hull-cluster bootstrap percentiles.}
\label{tab:iara}
\centering
\footnotesize
\begin{tabular}{lccccccc}
\hline
method & n & rank-1 & hit@5 & hit@10 & medR [95\% CI] & AUC [95\% CI] & EER \\
\hline
SKANN (published checkpoint) & 96 & 0.250 & 0.438 & 0.531 & 8 [3--15] & 0.824 [0.762--0.880] & 0.260 \\
narrowband-tonal & 96 & 0.260 & 0.458 & 0.510 & 10 [3--22] & 0.761 [0.691--0.827] & 0.333 \\
score fusion & 96 & 0.354 & 0.479 & 0.562 & 6 [2--17] & 0.839 [0.776--0.896] & 0.251 \\
\hline
\end{tabular}
\end{table*}

\emph{Paired contrasts (McNemar exact, discordant queries in favour of first / second method): fusion vs SKANN, hit@1 11/1, p = 0.006; hit@5 9/5, p = 0.42; hit@10 11/8, p = 0.65. SKANN vs tonal, hit@1 9/10, p = 1.0; hit@5 10/12, p = 0.83; hit@10 13/11, p = 0.84. Fusion vs tonal, hit@1 13/4, p = 0.049; hit@5 6/4, p = 0.75; hit@10 6/1, p = 0.13. Hull-bootstrap 95\% intervals on paired AUC differences: fusion $-$ SKANN [$-$0.025, +0.058]; SKANN $-$ tonal [$-$0.001, +0.126]; fusion $-$ tonal [+0.038, +0.118]. Nine contrasts, reported uncorrected.}

Three readings, stated at the resolution 96 queries from 37 hulls support.

\textbf{Cross-passage re-identification on public data is hard.} The correct hull tops the list in a quarter of queries for both methods, the genuine match sits at median rank 8 (embedding) or 10 (tonal) among 98 passage candidates, and even a ten-deep list contains it only about half the time. The hull-bootstrap intervals on median rank (3--15 and 3--22) show how much of that is sampling breadth: the gallery holds 40 hulls, and resampling them moves the median by a factor of two or more. These numbers support analyst triage over a ranked candidate list; they do not support treating either system as an identifier of record.

\textbf{The two methods are statistically indistinguishable at the top of the list.} Rank-1 differs by one query and hit@5 and hit@10 by two; the paired tests return p $\geq$ 0.83 at every depth, with discordant queries split almost evenly (9 vs 10 at rank-1). Where the methods separate is threshold-free ranking quality --- AUC 0.824 against 0.761, EER 0.260 against 0.333 --- but the paired AUC difference interval [$-$0.001, +0.126] reaches zero, so we report the direction as the observed pattern and not as an established ordering. Per-query genuine scores correlate at 0.55 between the methods: they agree on the easy and the hopeless queries and disagree on a band in the middle, which is what makes fusion worth reporting.

\textbf{Fusion improves the rank-1 point estimate, and that is the only contrast that reaches nominal significance.} Fusion converts 11 queries that the embedding alone missed at rank 1 and loses one (McNemar p = 0.006), taking rank-1 from 0.250 to 0.354. The gains at hit@5 and hit@10 (+2 and +3 queries) and in AUC (+0.015, interval spanning zero) are not distinguishable from noise, and with nine contrasts reported the rank-1 result sits at the boundary of a family-wise correction. Against the tonal comparator fusion is more clearly separated on AUC (+0.078, interval [+0.038, +0.118]). With 96 queries we present this as an observation about partial complementarity --- the methods fail on different queries in roughly a fifth of cases --- and not as an operational recommendation. We do not test how fusion behaves when a vessel is recorded at a different speed from its enrolment passage; \S\ref{sec:discussion} sets out why the public corpora cannot support that measurement.

\begin{figure}[!t]
\centering
\includegraphics[width=\columnwidth]{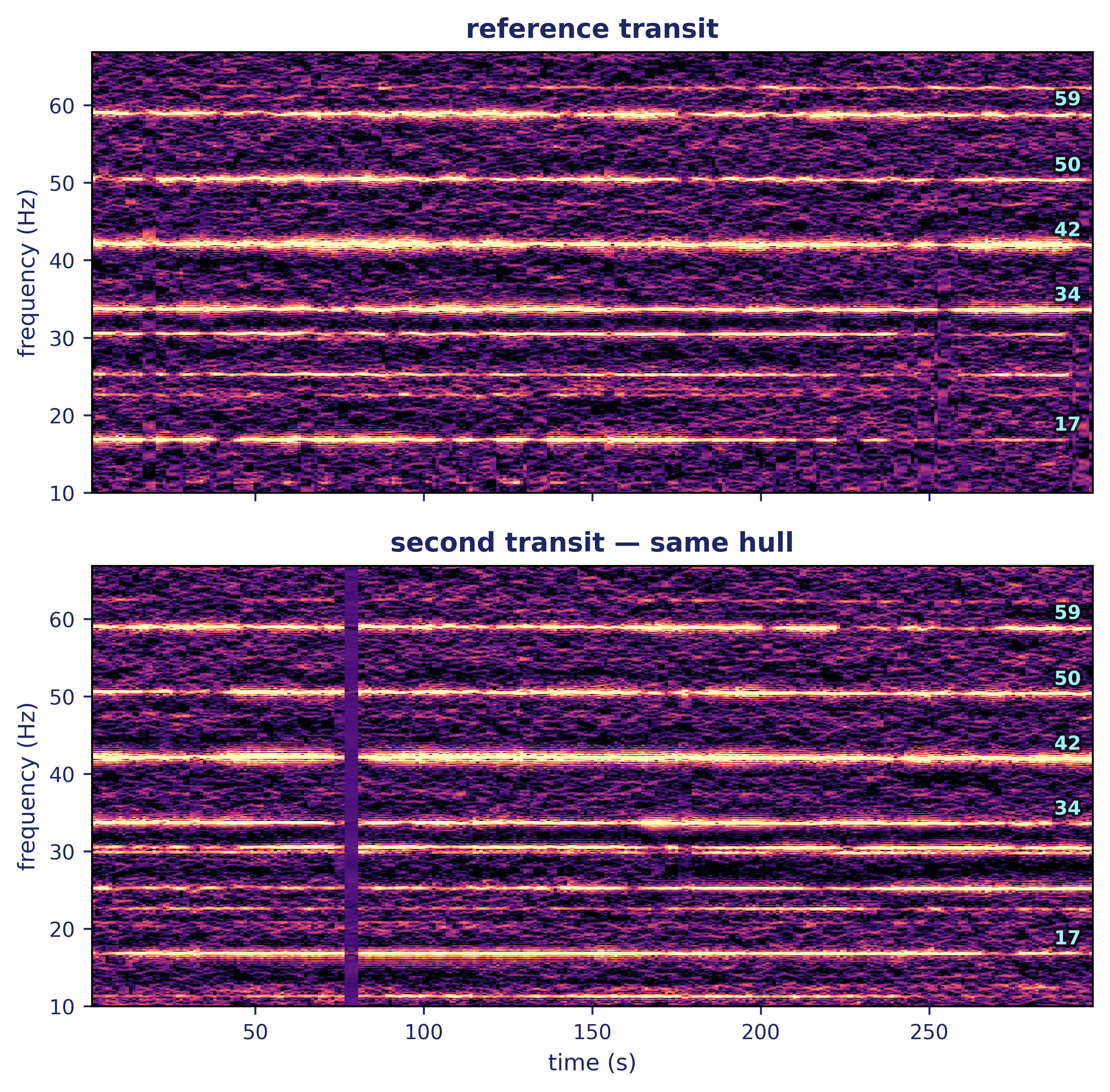}
\caption{Exemplar same-hull passage pair: whitened low-frequency spectrograms (10--65 Hz, 300 s) of a reference transit and a second, independent transit of the same hull from the IARA cross-passage evaluation. Persistent narrowband lines at 17, 34, 42, 50 and 59 Hz are common to both passages. Recording identifiers are withheld (\S\ref{sec:dedup}).}
\label{fig:pair}
\end{figure}

\begin{figure}[!t]
\centering
\includegraphics[width=0.95\columnwidth]{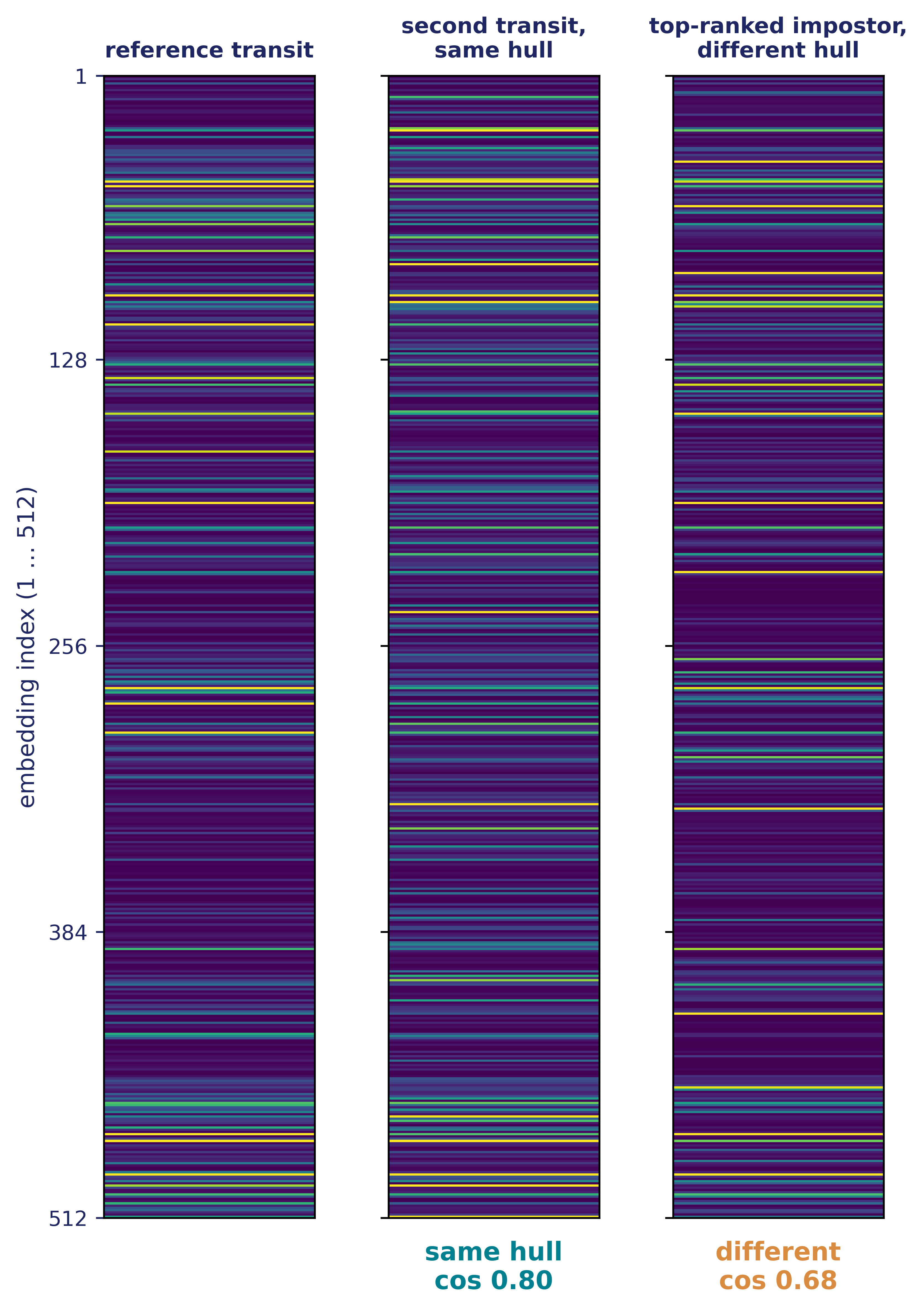}
\caption{Embedding view of the pair in Fig.~\ref{fig:pair}. The three 512-d SKANN embeddings are the two same-hull transits and the top-ranked different-hull impostor for the same query. Same-hull cosine similarity is 0.80; the impostor scores 0.68. The margin is representative of the top of the ranking, where correct matches and near-miss impostors are separated by tenths, not by a gap (\S\ref{sec:iara}).}
\label{fig:embed}
\end{figure}

\textbf{Effect of deduplication.} Scored identically without the transit map, the same checkpoint and harness return rank-1 0.409 / 0.470 / 0.539 over 115 queries --- a \textbf{16--21 point} apparent rank-1 advantage created by scoring recordings of the same physical transit against each other. This before/after contrast is paired on largely the same query set, and its direction is the finding: contamination masked weakness rather than manufacturing strength. We report both passes so readers can calibrate how much duplicate leakage can contribute to a headline number on this corpus --- and we note that any published recognition result treating IARA catalogue IDs as independent encounters is exposed to the same inflation. Figs.~\ref{fig:pair} and~\ref{fig:embed} illustrate one same-hull passage pair from this evaluation, as spectrograms and as embeddings.

\subsection{ShipsEar under an identity protocol}\label{sec:shipsear}

ShipsEar \cite{santosdominguez2016shipsear} is the most-used corpus in the underwater acoustic target recognition literature, and we included its nine multi-recording vessels in the hull-disjoint validation split (\S\ref{sec:data}). Under the cross-passage protocol of \S\ref{sec:problem} it does not function as a re-identification benchmark, and the manner of its failure is itself a finding.

\textbf{Supply.} The corpus contains 90 recordings of 58 named vessels, of which only 12 have more than one recording. Most of those repeats are state variants of a single docking event --- the vessel entering, waiting, and leaving, recorded minutes apart on the same day at the same site. They are genuine same-hull views at different speeds and aspects, but they are not independent passages: query and gallery share the day's water, the recording chain, and the ambient. The validation slice yields 14 scoreable queries against a 13-candidate source-pure gallery, drawn from \textbf{five} hulls with two or three passages each. The other four contribute nothing scoreable: each has two to four recordings, but all within a single session --- the recording manifest labels them \emph{arrives}, \emph{entering}, \emph{waiting}, \emph{leaves}, \emph{manoeuvre} --- so they collapse to one passage apiece under the session rule and cannot be queried at all. Thirty-seven validation recordings reduce to fourteen passages, a steeper collapse than IARA's 115 to 99. One query is $\approx$ 7 points of rank-1, and chance rank-1 on this gallery is $\approx$ 0.15.

\textbf{Result.} On this slice the embedding is at chance. Rank-1 is 0.143 (2 of 14, against a chance level of $\approx$ 0.15), the median rank of the genuine passage is 5.5 of 13, and the source-pure AUC is 0.560 (EER 0.432); the automated tonal method is likewise at chance (rank-1 0.000, AUC 0.500). More diagnostic than the headline is the geometry behind it. Across the 13 genuine pairs the mean same-vessel cosine is \emph{below} the mean different-vessel cosine (gap $-$0.066; pairwise EER 0.615), and the failed queries are not near-misses: the wrong vessel is returned at cosines of 0.80--0.92, values at which the same encoder returns correct matches on IARA. The embedding has not lost the signal in noise; it has merged the small craft of one harbour into a single tight region and then ranked them by whatever residual variation remains, which on this corpus is dominated by operating state rather than hull.

\textbf{The shortcut that hides it.} When the ShipsEar and IARA validation slices were first scored in one mixed gallery, ShipsEar AUC read 0.857. Scoring each corpus against its own gallery --- the source-pure rule of \S\ref{sec:crosspassage} --- returned it to 0.560. The 0.30 of AUC that vanished was the recording-chain signature: a ShipsEar genuine pair trivially out-scores an IARA impostor because both ShipsEar passages carry the same site, whereas on the honest within-corpus comparison that signature is common to every candidate and cancels. A mixed gallery therefore reports the encoder's ability to recognise \emph{where} a recording was made, and books it as identity.

\textbf{Reading.} We do not claim that ShipsEar is unusable for classification, for which it was built. We claim that its structure --- one site, one chain, repeats minutes apart --- cannot separate hull identity from channel and state under any protocol that scores it alone, and that any mixed-corpus retrieval number in which it participates should be read as at least partly a corpus-recognition number. A corpus can support cross-passage re-identification only if the same hull recurs across independent encounters; on this axis ShipsEar's 12 repeat vessels supply almost nothing, and the negative gap is the honest measurement of that. We report it because the alternative --- quoting the mixed-gallery 0.857 --- is the more common practice and is wrong.

\subsection{Cross-network fine-tuning: seen-vessel lift, unseen-vessel null}\label{sec:onc}

The published checkpoint is trained on IARA and ShipsEar only. ONC \cite{onc2020sog} is a separate hydrophone observatory with different hardware and deployment, and we used it to ask the transfer question directly: does a fingerprint learned on one recording network survive on another, and does fine-tuning on the new network help vessels the fine-tune never saw?

\textbf{Protocol.} The ONC multi-passage subset comprises 15 vessels with genuine cross-passage pairs. It was split before any fine-tuning into 8 \emph{enrolled} vessels, available for training, and 7 \emph{held-out} vessels, never trained on at any stage. Evaluation is cross-passage against a 97-passage ONC gallery (source-pure), reported on two query sets: the full 38 queries (which include the enrolled vessels' passages, and are therefore contaminated for any fine-tuned model) and the 17 held-out queries (the clean unseen-vessel verdict). Two fine-tunes were run from the published checkpoint: \textbf{A}, a small dose (the 8 enrolled vessels, 6 epochs), and \textbf{B}, a large dose (the 8 enrolled plus 82 single-passage ONC vessels added as new identities, 3 epochs), with the same 7 vessels held out from both. ONC has not been through the transit-level audit of \S\ref{sec:dedup}; its passages are the recording windows, with identities from AIS (\S\ref{sec:data}). One held-out query is $\approx$ 6 points of rank-1. Results are given in Table~\ref{tab:onc}.

\begin{table}[!t]
\caption{Cross-network fine-tuning on ONC: full (38-query) and held-out (17-query) sets against a 97-passage source-pure gallery.}
\label{tab:onc}
\centering
\footnotesize
\begin{tabular}{lccccc}
\hline
model & query set & n & rank-1 & medR & AUC \\
\hline
published (no ONC) & full-97 & 38 & 0.026 & 24.5 & 0.608 \\
fine-tune A (8 vessels) & full-97 & 38 & 0.105 & 13.0 & 0.682 \\
fine-tune B (90 vessels) & full-97 & 38 & 0.079 & 7.5 & 0.731 \\
tonal (automated) & full-97 & 38 & 0.053 & 26.0 & 0.639 \\
published (no ONC) & held-out & 17 & 0.059 & 53.0 & 0.527 \\
fine-tune A (8 vessels) & held-out & 17 & 0.000 & 41.0 & 0.516 \\
fine-tune B (90 vessels) & held-out & 17 & 0.000 & 36.0 & 0.551 \\
tonal (automated) & held-out & 17 & 0.000 & 20.0 & 0.599 \\
\hline
\end{tabular}
\end{table}

\textbf{Seen-vessel lift.} On the full query set every fine-tune improves every metric over the published model: median rank falls from 24.5 to 13.0 (A) and 7.5 (B), AUC rises from 0.608 to 0.682 and 0.731. The matched control makes the source of that lift unambiguous. The full set differs from the held-out set only by the 21 queries on the 8 enrolled vessels; the held-out rows are flat. The improvement is therefore located entirely on vessels the fine-tune trained on --- fine-tuning on locally labelled vessels roughly halved their median retrieval rank while leaving all other traffic where it was.

\textbf{Unseen-vessel null.} On the 7 held-out vessels neither dose moves the published model off chance. AUC 0.527 $\rightarrow$ 0.516 $\rightarrow$ 0.551 is within a few queries' worth of noise at n = 17 (a 95\% interval on any of these spans roughly 0.35--0.70); rank-1 is 1 of 17 or 0 of 17 throughout; median rank of the genuine passage stays in the thirties to fifties of 97. The automated tonal method is also at chance on this set (AUC 0.599, rank-1 0), so the null is not a property of the learned encoder alone: cross-recording re-identification on unseen hardware is not currently achievable by either method on this data. Increasing the fine-tuning dose tenfold in vessel count (8 $\rightarrow$ 90) did not change the outcome, which argues against the hypothesis that the small fine-tune was simply under-dosed.

\textbf{Diagnosis.} We attribute the null to the supply of cross-encounter positive pairs, not to the loss or the architecture. Of the 90 ONC vessels available for fine-tune B, 82 have a single recording. A single-recording identity can contribute positives only through augmentation of the one capture, which teaches the encoder that capture's invariances --- the channel, the ambient of that day --- rather than the invariance that transfers, which is the same hull under a genuinely different encounter. Only the 8 enrolled multi-passage vessels supplied the latter, and 8 hulls is not enough to move a 512-dimensional embedding's notion of what varies between encounters. This is the same constraint that shaped the base training (\S\ref{sec:data}, \S\ref{sec:epochset}): public corpora are singleton-heavy, and the quantity that determines cross-network generalisation is the count of hulls with independent repeat encounters, which no corpus we used supplies in volume on a second hydrophone network.

\textbf{Two caveats on the held-out numbers.} First, fine-tune B's checkpoint was selected on a validation gate that included the held-out ONC AUC among its inputs; its held-out number is therefore mildly optimistic by selection, which if anything strengthens the null. Second, the held-out set is seven vessels; repeated fine-tuning campaigns evaluated against the same seven would degrade it as a test set through selection, so we ran two fine-tunes and stopped. A sequestered third split would be required before iterating further.

\textbf{Reading.} Cross-network fine-tuning is an adaptation mechanism, not a generalisation mechanism, on the data presently available: it is expected to help on vessels for which the target network holds labelled repeat encounters, and there is no evidence that it helps on any other vessel. We report the unseen-vessel result as a null of equal standing with the seen-vessel lift, because the operational question --- will a fingerprint trained elsewhere recognise a new contact on my hydrophone --- is answered by the held-out rows, not the full rows.

\section{Discussion and limitations}\label{sec:discussion}

\subsection{What the results do and do not establish}\label{sec:establish}

The paper set out to measure open-set cross-passage re-identification on public data under a protocol that removes the two easiest ways to score well: seeing the query hull in training, and scoring a recording against another recording of the same transit. Under that protocol, three findings stand. First, the learned embedding and the automated tonal-line matcher are indistinguishable at rank 1 on a 40-hull gallery (0.250 vs 0.260 over 96 queries), and their score fusion, at 0.354, is the only contrast between the three systems that reaches nominal significance (\S\ref{sec:iara}). Second, the corpus audit matters more than the choice of method: removing same-transit duplicates lowers apparent rank-1 by 16--21 points on IARA, a larger movement than any between-method difference we measured (\S\ref{sec:dedup}, \S\ref{sec:iara}). Third, neither system generalises to a hydrophone network unseen in training (\S\ref{sec:onc}).

We therefore describe the system as a shortlisting instrument. An AUC of 0.82 and a median genuine rank of 8 on a 98-candidate list mean that an analyst handed the top ten will find the true hull in it slightly more often than not (hit@10 = 0.53), and that the learned encoder orders the remainder of the list more reliably than the tonal matcher does. Neither number licenses an identification: a rank-1 hit rate of one in four is a triage aid, not a verdict, and we have written the paper so that no sentence can be read as claiming otherwise.

The fusion result deserves a specific reading. The two systems' per-query genuine scores correlate at 0.55, so the fusion gain is real complementarity rather than averaging noise, and it is consistent with the two representations keying on different physical evidence --- absolute line frequencies in one case, a learned spectro-temporal texture in the other. But the gain is measured with a gallery-referenced normalisation (\S\ref{sec:metrics}) that assumes the gallery is representative of the score distribution at deployment; a gallery of different composition would shift the normalisation and could shift the gain. We report fusion as evidence that the representations are partly complementary, not as a recommended operating configuration.

\subsection{Limitations}\label{sec:limitations}

We order these by consequence for a reader deciding whether to build on the result.

\textbf{Cross-network generalisation is a measured null.} The strongest negative finding in the paper is that a hull fingerprint learned on one hydrophone network does not recognise vessels on another (\S\ref{sec:onc}). On seven vessels held out from fine-tuning, both the learned encoder and the tonal matcher stayed at chance, and increasing the fine-tuning dose tenfold in vessel count did not move them. Fine-tuning on the target network helped only the vessels it was trained on. Every headline number in \S\ref{sec:iara} is therefore a within-network number: the IARA gallery and the IARA queries were captured on the same hardware, and the paper offers no evidence that its ranking quality survives a change of hydrophone, mooring, or front-end electronics. For the operational question that motivates this work --- will a fingerprint enrolled elsewhere recognise a new contact on my sensor --- the honest current answer is no.

\textbf{Speed variation is untested.} A vessel's radiated tonal set shifts with shaft and blade rate, so a passage recorded at a different speed from the enrolment passage is the obvious stress case for any method that keys on absolute line frequencies --- and, by extension, for a fusion that includes one \cite{kim2025vessel}. We do not report a controlled measurement of this. Synthetic resampling would not provide one: it displaces line frequencies without reproducing the accompanying changes in source level, cavitation onset, and modulation structure, so it tests robustness to an artefact rather than to the physical condition. A genuine measurement requires a corpus with speed-over-ground logged per passage and the same hull recorded across a speed range; none of the three public corpora used here provides it. We regard this as the most consequential gap in the evaluation and the clearest requirement for future data collection.

\textbf{The evaluation resolves coarse differences only.} Ninety-six queries drawn from 37 hulls is a small instrument. One query is about one percentage point of rank-1, and because queries from the same hull are not independent, 37 hulls --- not 96 queries --- is the effective sample size for any between-method claim; our intervals are hull-cluster bootstraps for that reason (\S\ref{sec:metrics}, \S\ref{sec:iara}). Nine paired contrasts are reported without correction for multiplicity, and none survives a Bonferroni adjustment at the 5\% level --- the fusion rank-1 result, the strongest of the nine, has an exact McNemar p of 0.006 against an adjusted threshold of 0.0056. We call it nominally significant for that reason and no stronger, and we would not defend the AUC ordering between the learned and tonal systems as established, since its interval reaches zero. Readers should treat the between-method table as a set of directions consistent with the data, and the protocol findings (deduplication effect, unseen-hardware null) as the results with margin to spare.

\textbf{Cross-encounter supply is the binding constraint on training.} The identity set used to train the encoder is singleton-heavy: 500 of 646 training hulls contribute a single recording (\S\ref{sec:data}, \S\ref{sec:training}). A singleton can supply positive pairs only through augmentation of one capture, which teaches the encoder that capture's channel and ambient rather than the invariance that transfers --- the same hull under an independent encounter. The same constraint reappears in the cross-network experiment, where 82 of 90 target-network vessels were singletons (\S\ref{sec:onc}). We attribute both the modest absolute performance and the cross-network null to this supply rather than to the loss, the front end, or the architecture, and we note that the quantity a future corpus should maximise is not hours or vessel count but \emph{hulls with repeat encounters on independent days and, ideally, independent sensors}.

\textbf{The most-used public corpus is not an identity benchmark.} Under an identity protocol, ShipsEar ranks by operating state rather than hull: mean same-vessel similarity falls below mean different-vessel similarity on its single recording chain (\S\ref{sec:shipsear}). This is not a criticism of the corpus for the purpose it was built for. It is a caution that results labelled ``re-identification'' on ShipsEar, or on any mixed gallery that includes it, should be read as partly corpus- and state-recognition until the protocol is shown to control for both.

\textbf{Single-sensor, single-channel assumptions.} Every experiment here scores one hydrophone channel against a gallery captured on the same class of sensor. We make no claim about arrays, beamformed inputs, multi-sensor fusion, or towed systems, and the gallery-referenced score normalisation used for fusion is defined relative to a specific gallery and would need to be re-derived for another.

\textbf{The duplicate-transit audit is corpus-specific.} The deduplication rule of \S\ref{sec:dedup} was calibrated on IARA's metadata and waveform structure. Other corpora may lack the closest-point-of-approach and timing fields it relies on, and the thresholds are unlikely to transfer unchanged. What transfers is the requirement: an identity protocol must state how it established that query and gallery are independent captures, and we would regard any cross-passage number that does not as unverified.

\textbf{No sea-trial or operational validation is claimed.} All results are retrospective evaluations on archived public recordings. The accept/reject threshold that an open-set system needs at deployment (\S\ref{sec:reid}) has not been calibrated against an operational false-alarm budget, and we have not measured analyst-in-the-loop performance. The comparison in this paper is automated-against-automated throughout.

\textbf{The held-out cross-network split should not be reused.} Seven vessels is a small held-out set, and repeated fine-tuning campaigns evaluated against it would erode it through selection. We ran two fine-tunes and stopped. A sequestered split would be required before further iteration on that network, and we recommend that any group extending \S\ref{sec:onc} constitute one before they begin.

\subsection{What would change the picture}\label{sec:change}

The limitations above point to a short list of data rather than method requirements. A corpus with (i) repeat encounters of the same hull on independent days, (ii) the same hulls captured on at least two independent sensor installations, and (iii) speed-over-ground logged per passage would let the three open questions of this section --- cross-network generalisation, speed robustness, and the reliability of the fusion gain --- be answered with the protocol already defined here. We release the protocol, the deduplication audit, the weights and the per-query outputs \cite{tyagi2026skann} so that such a corpus, when it exists, can be evaluated against the same instrument.

\section{Conclusion}\label{sec:conclusion}

We have specified a protocol for open-set, cross-passage vessel re-identification on public hydrophone data --- hull-disjoint splits keyed to vessel identity, galleries and queries drawn from disjoint passages, source-pure galleries, and an audio-adjudicated transit-deduplication gate --- and evaluated under it a raw-waveform selective-kernel encoder, a fully automated narrowband-tonal comparator, and their score fusion. The headline finding is not a method ranking. On a 40-hull IARA gallery the embedding and the tonal matcher are indistinguishable at the top of the list (rank-1 0.250 vs 0.260 over 96 queries); the embedding orders the remainder more reliably (AUC 0.824 vs 0.761) with an interval that reaches zero; and fusion lifts rank-1 to 0.354, the one contrast that reaches nominal significance and none that survives a family-wise correction. Two protocol findings carry more weight than any of these: removing same-transit duplicates lowers apparent rank-1 by 16--21 points, more than any between-method difference we measured; and neither system recognises vessels on a hydrophone network unseen in training. ShipsEar, the field's most-used corpus, cannot separate hull identity from recording channel under an identity protocol.

The practical reading is that current public data supports analyst triage over a ranked shortlist, and does not support treating any of the three systems as an identifier of record. The learned front end nevertheless tells us something about the problem: from random initialisation it reconstructs a constant-Q decomposition tuned below 500 Hz together with harmonic-comb detectors on its longest branch, which is the physics of ship-radiated noise recovered from the task rather than imposed on it.

What would change the picture is data, not architecture: a corpus with repeat encounters of the same hull on independent days, the same hulls captured on at least two independent sensor installations, and speed over ground logged per passage. The protocol, the deduplication audit, the weights and the per-query outputs are released so that such a corpus, when it exists, can be evaluated against the same instrument.

\section*{Data and Code Availability}

\textbf{Corpora.} IARA \cite{dasilva2025iara} was used at Zenodo record doi:10.5281/zenodo.15758636, which is byte-identical in all audio archives and the metadata spreadsheet to record 15777429, the record the IARA paper cites (the two differ only in a summary figure); the concept DOI is 10.5281/zenodo.15758635. ShipsEar \cite{santosdominguez2016shipsear} was used under its authors' terms. Ocean Networks Canada hydrophone recordings (SCVIP, 1--27 July 2020) and the AIS logs used to key them to vessels (Iona Island receiver, 1--31 July 2020) were obtained through the Oceans 3.0 portal and are cited per dataset \cite{onc2020sog}. No audio or derived tensors from any corpus are redistributed.

\textbf{Artefacts.} The published SKANN checkpoint (best-validation epoch 21), the validation passage embeddings re-extracted from it, the audio-adjudicated transit map of \S\ref{sec:dedup} together with the script that produced it, the validation similarity matrices for both methods, and the per-query results behind \S\ref{sec:iara} are deposited under CC-BY-4.0 at doi:10.5281/zenodo.22160138 \cite{tyagi2026skann} (concept DOI 10.5281/zenodo.22160137), with file checksums in the deposit README. Re-extracting the passage embeddings from the deposited checkpoint reproduces every reported similarity to within $10^{-4}$, and every table in \S\ref{sec:iara} can be regenerated from the deposited similarity matrices without audio or model. The deposit contains no audio.

\textbf{Comparator.} The narrowband-tonal comparator is fully specified in \S\ref{sec:metrics} (TPSW whitening, persistent-line extraction, matched-line-fraction scoring); no analyst-in-the-loop numbers are reported.

\end{document}